\documentclass{emulateapj}
\usepackage{bm}

\newcommand{\gcc}{\mbox{ g cm$^{-3}$}}
\newcommand{\beq}{\begin{equation}}
\newcommand{\eeq}{\end{equation}}
\newcommand{\bea}{\begin{eqnarray}}
\newcommand{\eea}{\end{eqnarray}}
\newcommand{\req}[1]{Eq.\ (\ref{#1})}
\newcommand{\mel}{m_\mathrm{e}}

\newcommand{\am}{a_\mathrm{m}}
\newcommand{\omc}{\omega_\mathrm{c}}
\newcommand{\omg}{\omega_\mathrm{g}}

\newcommand{\kB}{k_\mathrm{B}}

\newcommand{\rb}[1]{\raisebox{1.5ex}[0pt]{#1}}

\shorttitle{Thermal Structure and Cooling of Superfluid Magnetic Neutron Stars}
\shortauthors{A. Y. Potekhin, D. G. Yakovlev, G. Chabrier, \& O. Y. Gnedin}

\begin{document}

\title{Thermal Structure and Cooling 
of Superfluid Neutron Stars
with Accreted Magnetized Envelopes}

\author{Alexander Y. Potekhin,\altaffilmark{1,2} 
Dmitry G. Yakovlev,\altaffilmark{1}
Gilles Chabrier,\altaffilmark{3}
 and Oleg Y. Gnedin\altaffilmark{4}}
 
\slugcomment{Received 2003 February 21; accepted 2003 May 13}

\altaffiltext{1}{Ioffe Physico-Technical Institute,
    Politekhnicheskaya 26, 194021 St.~Petersburg, Russia:
    palex@astro.ioffe.ru, yak@astro.ioffe.ru}
\altaffiltext{2}{Isaac Newton Institute of Chile, St.~Petersburg Branch, Russia}
\altaffiltext{3}{Ecole Normale Sup\'erieure de Lyon
    (C.R.A.L., UMR CNRS No.\ 5574),
    46 all\'ee d'Italie, 69364 Lyon Cedex 07, France:
    chabrier@ens-lyon.fr}
\altaffiltext{4}{Space Telescope Science Institute,
         3700 San Martin Drive, Baltimore, MD 21218, USA:
         ognedin@stsci.edu}

\begin{abstract}
We study the thermal structure of neutron stars with magnetized envelopes
composed of accreted material,
using updated thermal conductivities of plasmas in quantizing magnetic fields,
as well as equation of state and
radiative opacities for partially ionized hydrogen 
in strong magnetic fields.
The relation between the internal
and local surface temperatures is calculated
and fitted by an analytic function
of the internal temperature, magnetic field strength,
angle between the field lines and the normal to the surface,
surface gravity, and the mass of the accreted material.
The luminosity of a neutron star with a dipole magnetic field
is calculated for various values of
the accreted mass, internal temperature, and magnetic field strength.
Using these results, we simulate cooling
of superfluid neutron stars with magnetized accreted envelopes.
We consider slow and fast cooling regimes, paying special attention to
very slow cooling of low-mass superfluid neutron stars.
In the latter case, the cooling is strongly affected by the combined 
effect of magnetized accreted envelopes and neutron superfluidity
in the stellar crust. Our results are important for interpretation of
observations of isolated neutron stars hottest for their age,
such as RX J0822$-$43 and PSR B1055$-$52.
\end{abstract}
\keywords{dense matter---magnetic fields---stars: 
individual (RX J0822.0$-$4300, PSR B1055$-$52)---stars: neutron}
\slugcomment{The Astrophysical Journal, 594: 404--418, 2003 September 1}
\maketitle

\section{Introduction}
Cooling of neutron stars (NSs) depends on the properties 
of dense matter in their crusts and cores. 
Theoretical predictions of these properties
depend on a model of strong interaction
and a many-body theory employed.
Therefore, one can test microscopic models
of dense matter by comparing
the results of simulations of NS cooling with observations
of the stellar thermal emission
(e.g., \citealt{pethick,page97,page98}).
The history of NS cooling theory is reviewed, e.g.,
by \citet*{yls99}; the main cooling regulators
are discussed, e.g., by \citet{badhonnef}.
Practical implementation of this method is restricted by
the accuracy of observational data 
and by the quality of the physics input
used in calculations. 

It is conventional (e.g., \citealt*{GPE})
to separate the outer
\emph{heat-blanketing envelope} 
from the stellar interior. The  
blanketing envelope can be treated separately
in the plane-parallel quasi-Newtonian approximation
\citep{GPE}, which simplifies the cooling calculations.

The NS cooling is
strongly affected by the relation between the heat flux density $F$
 and the
 temperature at the inner boundary of the blanketing envelope, 
 $T_\mathrm{b}$. The flux density
 is constant throughout a given local part of the envelope. It
 is conventionally written as $F=\sigma T_\mathrm{s}^4$,
 where $T_\mathrm{s}$ is the local effective surface temperature,
 and $\sigma$ is the Stefan--Boltzmann constant.
The $T_\mathrm{b}$--$T_\mathrm{s}$ relation is
determined by the equation of state (EOS) and thermal conductivity of
matter in the heat-blanketing envelope.

For nonmagnetized stellar envelopes composed of iron,
this relationship was thoroughly studied
by \citet{GPE}. \citet*{CPY} and \citet*{PCY} (hereafter Paper~I), 
reconsidered the problem and 
 extended the results 
in two aspects. First, advanced
theoretical data on EOS and thermal
conductivity allowed the authors to study
colder NSs, with $T_\mathrm{s}$ down to $5\times10^4$~K.
Second, the authors considered the blanketing
envelopes composed not only of iron but also of
light (accreted) elements.
They found that a small amount of accreted matter
(with mass $10^{-16}\, M_\odot\lesssim\Delta M 
\lesssim 10^{-7}\, M_\odot$) result in higher $T_\mathrm{s}$
at a given $T_\mathrm{b}$. As a consequence,
the stellar luminosity is higher at the 
early (neutrino-dominated) cooling stage 
(at stellar ages $t \la 10^4 - 10^5$~yr)
and lower at the subsequent photon stage.

Recently the thermal structure of accreted envelopes
has been studied by \citet*{bbc02} taking into account 
that hydrogen burning in accreted matter may proceed
\citep{schatzetal01}
far beyond Fe (up to Te) via the rapid proton
capture process.

It was realized long ago \citep{Tsuruta72} that strong magnetic
fields can greatly affect the surface temperature and cooling  of
NSs. The thermal structure of NS envelopes with radial magnetic
fields (normal to the surface) was studied by \citet{kvr88} (also see
\citealt{kvr88} for references to the earlier work). His principal
conclusion was that the field reduces the thermal insulation of the
blanketing envelope due to Landau quantization of electron motion.

The thermal structure of the envelope with magnetic fields
normal and tangential to the surface was analyzed by
\citet{hern85} and \citet{schaaf90a}.
The tangential field increases the thermal insulation
of the envelope, because the Larmor rotation of the electrons
reduces
the transverse electron thermal conductivity.

The case of arbitrary angle $\theta$ between the field lines
and the normal to 
the surface was studied by \citet{gh83} in the approximation
of constant (density and temperature independent)
longitudinal and transverse thermal conductivities.
The authors proposed a simple formula
which expresses $T_\mathrm{s}$ at arbitrary $\theta$
through two values of  $T_\mathrm{s}$ 
calculated at $\theta=0$ and $90^\circ$.
The case of arbitrary $\theta$
was studied also by \citet{schaaf} 
and \citet{hh-theory,hh-multi}.

\citet{PY01} (hereafter Paper~II) reconsidered the thermal structure
of the blanketing envelopes for any $\theta$, using improved thermal
conductivities \citep{P99}. In agreement with an earlier conjecture
of \citet{hern85} and simplified treatments of \citet{page95} and
\citet{shibyak}, Paper~II demonstrated that the dipole magnetic field
(unlike the radial one) does not necessarily increase the total
stellar luminosity $L_\gamma$ at a given $T_\mathrm{b}$. On the
contrary, the field $B\sim10^{11}$--$10^{13}$~G lowers $L_\gamma$,
and only the fields $B\gtrsim10^{14}$ G significantly increase it.

Early simulations of cooling of magnetized NSs
were performed assuming the radial magnetic field
everywhere over the stellar surface
(e.g., \citealt{Tsuruta72,nt87,kvr91}).
Since the radial magnetic field reduces the thermal insulation,
these models predicted the increase of  $L_\gamma$
by the magnetic field at the neutrino cooling stage.
\citet{page95} and \citet{shibyak} simulated cooling
of NSs with dipole magnetic fields, using the Greenstein--Hartke 
formula and the $T_\mathrm{b}$--$T_\mathrm{s}$ relations
at the magnetic pole and equator 
taken from the previous work \citep{hern85,kvr88,schaaf90a}.
\citet{HH97,HH97b,hh-theory,hh-multi} 
proposed simplified
models of cooling magnetized NSs including the cases
of ultrahigh surface magnetic fields, $B \sim 10^{15}$--$10^{16}$ G.
Recently, the improved $T_\mathrm{b}$--$T_\mathrm{s}$ relation derived 
in Paper~II has been used for simulating the
cooling of NSs with dipole magnetic fields
(Paper~II; \citealt*{kyg02,badhonnef}).

Most of the above-cited work on magnetized NSs has been focused on
NSs with iron envelopes, except that \citet{HH97} have used
simplified analytic models to calculate cooling of ultramagnetized
($B\sim10^{15}$--$10^{16}$~G) NSs which may possess accreted
envelopes.

In this paper, we consider accreted NS envelopes  (as in Paper~I for
$B=0$), but take into account strong magnetic fields $B$ directed at
various angles $\theta$ (as in Paper~II for iron envelopes). We use
the modern electron conductivities of magnetized  envelopes
\citep{P99}, taking into account the effects of finite size of atomic
nuclei, appropriate for the inner envelopes \citep*{gyp},  as well as
the radiative opacities and EOS of strongly magnetized, partially
ionized hydrogen in NS atmosphere \citep{PC02}. We derive an analytic
approximation for the surface temperature. We calculate
$T_\mathrm{eff}$ for a dipole magnetic field and simulate NS cooling
to demonstrate the effects of magnetic fields and accreted envelopes
on thermal evolution of NSs.

In \S\ref{sect-mf-effects} we give an overview of the main effects of
strong magnetic fields on heat conduction in a plasma, and introduce
basic definitions. In \S\ref{sect-input} we describe the physics input
used in our simulations. The thermal structure of NS envelopes with
magnetic fields is discussed in \S\ref{sect-thst}. NS cooling
simulations are presented in \S\ref{sect-cool}.
The conclusions are formulated
in \S\ref{sect-concl}.

\section{The effects of strong 
magnetic fields on heat conduction in neutron star envelopes}
\label{sect-mf-effects}

We call a magnetic field \emph{strong}\
if the electron cyclotron energy $\hbar\omc=\hbar eB/\mel c$
exceeds 1 a.u.\ --- i.e., the field strength $B$
is higher than $\mel^2 c\, e^3/\hbar^3
= 2.3505\times10^9$~G,
where $\mel$ is the electron mass, $e$ the elementary charge,
and $c$ the speed of light. 
We call the field \emph{superstrong}
if $\hbar\omc > \mel c^2$,
that is $B > \mel^2 c^3/e\hbar
= 4.414\times10^{13}$~G.

The field is called \emph{strongly quantizing} if it sets
almost all electrons on the ground Landau level. 
The latter case takes place 
at temperature $T\ll T_B$ and density $\rho<\rho_B$, where
\begin{eqnarray}
   \rho_B & = & m_\mathrm{u} n_B \, \langle A\rangle/\langle Z\rangle
  \approx 7045 \,B_{12}^{3/2}\,(\langle A\rangle/\langle Z\rangle)\gcc,
\label{rho_B}
\\
    T_B & = & {\hbar\omc/\kB} 
    \approx 1.343\times10^8\, B_{12}{\rm~K}.
\label{T_B}
\end{eqnarray}
Here, $m_\mathrm{u} = 1.66054\times10^{-24}$~g is the atomic mass unit,
$\langle A\rangle$ and $\langle Z\rangle$ 
are the mean ion mass and charge numbers, $B_{12}\equiv B/(10^{12}$~G),
$n_B=1/(\pi^2\sqrt2\,\am^3)$ is the electron number density
at which the electron 
Fermi energy reaches the first excited Landau level,
and $\am=(\hbar c/eB)^{1/2}$ is the quantum magnetic length.

As a rule, at $\rho > \rho_B$ and not too high $T$, 
the electron thermodynamic and 
kinetic functions oscillate, with increasing $\rho$,
around their values calculated neglecting quantization
of electron motion in Landau levels.
If $T \gtrsim T_B$ or $\rho\gg\rho_B$,
the field can be treated as classical (\emph{nonquantizing}).

The increase of the atomic binding energies in the strong magnetic field
tends to lower the ionization degree, as first guessed by
\citet*{CLR70}. Hence, there can be a significant
amount of bound species in a highly magnetized atmosphere,
even if it were negligibly small at the same temperature
in the zero-field case. This idea has recently been confirmed
 by detailed NS atmosphere models \citep*{RRM,PCS99,PC02}.

The effects of magnetic fields on the kinetic properties 
of the outer NS layers have been reviewed, e.g., by
\citet{YaK} and \citet{elounda}.
A strongly quantizing magnetic field reduces the 
mean radiative opacities 
(e.g., \citealt{silyak}); therefore 
the bottom of the photosphere shifts to higher densities 
(e.g., \citealt{Pavlov95,LS97}).

A nonquantizing magnetic field does not affect 
thermodynamic functions of the plasma. However,
it does affect the electron heat conduction, unless
the \emph{magnetization parameter} 
\beq
   \omg\tau\approx 1760\,\frac{B_{12}}{\sqrt{1+x_\mathrm{r}^2}}
   \,\frac{\tau}{10^{-16}\mathrm{~s}}
\label{omegatau}
\eeq
 is small.
Here, $\omg=\omc/\sqrt{1+x_\mathrm{r}^2}$ 
is the electron gyrofrequency,
\begin{equation}
   x_\mathrm{r} = {{\hbar(3\pi^2n_\mathrm{e})^{1/3}}/({\mel c}})
   \approx
   1.0088 \,(\rho_6\,Z/ A )^{1/3}
\label{x_r}
\end{equation}
is the \emph{relativity parameter}, 
$\rho_6\equiv\rho/(10^6\gcc)$, and $\tau$ is an effective relaxation
time. In a degenerate Coulomb plasma with a nonquantizing
magnetic field,
\beq
   \tau \approx  { 3 \pi \hbar^3 \over 
   4 Z \mel e^4 \Lambda\,\sqrt{1+x_\mathrm{r}^2}}
   =\frac{5.700\times10^{-17}}{Z\Lambda\,\sqrt{1+x_\mathrm{r}^2}}
   \mbox{~s},
\label{tau}
\eeq
where $\Lambda$ is an effective Coulomb logarithm.
One has $\Lambda\sim1$ in the ion liquid, and $\Lambda$ is roughly
proportional to $T$ in the classical Coulomb solid
(e.g., \citealt{YU}).
However, according to \citet{baiko-ea98},
the correlation effects smooth
the dependence of $\Lambda$ on $T$ or $\rho$
near the melting point and introduce deviations from
this proportionality.

In a magnetic field, one should generally introduce
two different relaxation times ($\tau_\|$ and $\tau_\perp$),
determined by two Coulomb logarithms ($\Lambda_\|$ and $\Lambda_\perp$),
corresponding to electron transport parallel and perpendicular to 
the field lines. Analytic fits for $\Lambda_\|$ and $\Lambda_\perp$
at any $B$ have been constructed by \citet{P99}.

As seen from Eqs.\ (\ref{omegatau}) and (\ref{tau}), the
magnetization parameter is large in the outer NS envelope at
$B\gtrsim10^{11}$~G, typical for many
isolated NSs. Moreover, according to Eqs.\ (\ref{rho_B})
and (\ref{T_B}) the magnetic field can be strongly quantizing
in the outermost part of the envelope.
Therefore, the field can greatly affect the heat conduction 
and the thermal structure 
of the NS envelope.
In this case, the surface temperature $T_\mathrm{s}$
may be nonuniform, depending on the magnetic field geometry.
Then, we introduce the effective temperature of the star
$T_\mathrm{eff}$ defined by
\begin{equation}
   4 \pi \sigma R^2 T_\mathrm{eff}^4 =
   L_\gamma= 
 \int F \,\mathrm{d}\Sigma =
\sigma \int T_\mathrm{s}^4 \,\mathrm{d}\Sigma
  ,
\label{L}
\end{equation}
where $R$ is the stellar (circumferential) radius, and
 $\mathrm{d}\Sigma$ is the surface element.
The quantities $T_\mathrm{s}$, $T_\mathrm{eff}$, and $L_\gamma$
refer to a local reference frame at the NS surface.
The redshifted (``apparent'') quantities
as detected by a distant observer are \citep{thorne}:
$T_\mathrm{s}^\infty = T_\mathrm{s} \, \sqrt{1-r_g/R}$,
$T_\mathrm{eff}^\infty = T_\mathrm{eff} \, \sqrt{1-r_g/R}$,
and
$L_\gamma^\infty = (1-r_g/R)\, L_\gamma$,
where 
$r_g=2GM/c^2=2.95(M/M_\odot)$ km is the Schwarzschild radius
defined by the total gravitational NS mass $M$, and
$G$ is the gravitational constant.

\section{Physics Input}
\label{sect-input}

Much of the physics input used in our work is 
the same as in Papers~I and II. 
We outline these cases and
refer the reader to
Papers~I and II for details.

\subsection{Basic Equations} 
\label{sect-equations} 

We choose the boundary between the internal NS region  and the
heat-blanketing envelope at the neutron-drip density 
$\rho_\mathrm{b}=4\times10^{11}\gcc$ (at a radius $R_\mathrm{b}$
a few hundred meters under the surface).
Although $\rho_\mathrm{b}=10^{10}\gcc$ adopted by \citet{GPE} ensures
the necessary condition $R-R_\mathrm{b}\ll R$ with higher accuracy,
the present choice better conforms to the requirement that $T$ does
not vary over the  boundary, if a strong magnetic field is present
(Paper~II). In addition, we show that this choice provides a better
accuracy to the $T_\mathrm{s}$--$T_\mathrm{b}$ relation in the
presence of an accreted envelope.
On the other hand, the increase of
$\rho_\mathrm{b}$ increases the thermal relaxation time of the
blanketing envelope, thus hampering a study of rapid variability of
the thermal emission (over time scales $\lesssim10$ yr
for $\rho_\mathrm{b} \sim 4\times10^{11}\gcc$,
\citealt{ur01})
with a cooling code.

The thermal structure of the blanketing envelope is studied
in the stationary, local plane-parallel approximation,
assuming that a scale of
temperature variation over the surface is much larger than the
thickness of the blanketing envelope.
This leads to the \emph{one-dimensional} approximation
for the heat diffusion equation:
\begin{equation}
 F  = \kappa {{\rm\,d}T\over{\rm\,d}z} ,
\quad
    \kappa\equiv{16\sigma T^3\over3K\rho},
\label{Flx}
\end{equation}      
where $\kappa$ is an effective thermal conductivity
along the normal to the surface, $K$ is the mean opacity,
and $z=(R-r)\,(1-r_g/R)^{-1/2}\ll R$ is the local proper depth
in the envelope.

Equation (\ref{Flx}) can be reduced (e.g., \citealt{kvr88}) to
\begin{equation}
   {\mathrm{d}\log T\over\mathrm{d}\log P} =
   {3\over 16}\,{PK\over g}\,{T_\mathrm{s}^4\over T^4},
\label{th-str}
\end{equation}
where $P$ is the pressure.
We integrate \req{th-str} inwards from the radiative surface, 
where $T=T_\mathrm{s}$, 
using the same algorithm as in Paper~I. We place 
the radiative surface at the Rosseland optical depth equal 
to $2/3$.\footnote{%
Assuming the local thermodynamic 
equilibrium, this depth
 follows from the Milne--Eddington formula, which is fairly accurate
as shown by \citet{Kourganoff}.}
Another boundary condition would imply another integration
constant for the integral of \req{th-str},
which becomes negligible in the inner part of the envelope,
where $T\gg T_\mathrm{s}$ (e.g., \citealt{elounda}). 
Therefore the choice of the boundary
condition does not noticeably affect the heat flux and the 
$T_\mathrm{s}$--$T_\mathrm{b}$ relation.

The validity of the one-dimensional approximation can be checked by
two-dimen\-sion\-al simulation of the heat transport  in the
blanketing envelope.  Such a simulation was attempted by
\citet{schaaf} for a homogeneously magnetized NS under many
simplified assumptions. The heat conduction from hotter to cooler
zones along the surface  or possible meridional and convective
motions can smooth the temperature variations over the NS surface.
Nevertheless, the one-dimensional approximation seems to be
sufficient to simulate NS cooling.

The thermal conductivity tensor of a magnetized plasma is anisotropic.
It is characterized by the conductivities parallel ($\kappa_\|$)
and perpendicular ($\kappa_\perp$) to the field, and by the 
off-diagonal (Hall) component.
In the plane-parallel approximation, Eq.~(\ref{Flx})
contains the effective thermal conductivity
\begin{equation}
   \kappa=\kappa_\|\cos^2\theta + \kappa_\perp\sin^2\theta.
\end{equation}

Assuming the dipole field, we
 use the general-relativistic formulas
\citep{GinzOz}
\begin{mathletters}
\begin{eqnarray}
&&
  B(\chi) = B_\mathrm{p}\,\sqrt{\cos^2\chi+a^2\sin^2\chi},
  \label{dipole}
\\&&
  \mathrm{tan}\,\theta = a\,\mathrm{tan}\,\chi,
\label{theta-chi}
\\&&
  a = - {(1-x) \ln(1-x) + x - 0.5 x^2
        \over [ \ln(1-x)+x + 0.5 x^2] \sqrt{1-x}},\hspace*{2em}
\end{eqnarray}
\end{mathletters}

\noindent
where $B_\mathrm{p}$ is the field strength at the magnetic pole,
$\chi$ is the polar angle, and $x=r_g/R$.

\subsection{Equation of state}
\begin{figure}\epsscale{1}
\plotone{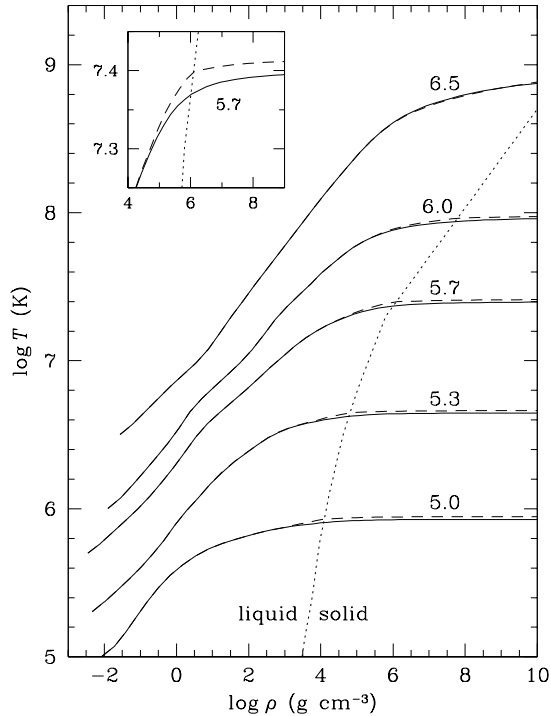}
\caption{Temperature profiles in a nonaccreted, nonmagnetized
envelope of a ``canonical'' NS ($M=1.4\,M_\odot$, $R=10$ km)
for several values of $T_\mathrm{s}$ (marked by $\log T_\mathrm{s}$).
Solid lines: calculation with the improved electron conductivities;
dashed lines: Paper~I. The dotted line shows the liquid/solid
boundary. Inset shows a temperature profile near this boundary
in more detail.
\label{fig-ps2fe}}
\end{figure}

A nonaccreted heat-blanketing envelope is assumed to be composed of iron,
which can be partially ionized at $\rho\lesssim10^6\gcc$.
Ions in the envelope form either Coulomb liquid or Coulomb crystal.
The melting of the crystal occurs 
at $\Gamma=\Gamma_\mathrm{m}\approx175$ (e.g., \citealt{PC00}),
where $\Gamma=(Ze)^2/\kB T a_\mathrm{i}$ 
is the ion coupling parameter,
 $a_\mathrm{i}=(4\pi n_\mathrm{i}/3)^{-1/3}$ is the ion sphere radius,
and $n_\mathrm{i}$ is the ion number density.
The melting curve $\Gamma=\Gamma_\mathrm{m}$ is shown 
in Fig.~\ref{fig-ps2fe} by the dotted line. 
The figure demonstrates
the thermal structure of a nonmagnetic iron envelope.
The value of the surface gravity $g=2.43\times10^{14}$ cm s$^{-2}$
chosen in this example corresponds to the ``canonical'' NS model
 with $M=1.4\,M_\odot$ and $R=10$ km.
At $B=0$, as in Paper~I, we use 
the OPAL EOS \citep*{OPAL-EOS}, extrapolated beyond 
the available tables
using an effective charge number $Z_\mathrm{eff}$.
In the strong magnetic field we employ
the finite-temperature Thomas--Fermi EOS of 
\citet{Thorolfsson},\footnote{Available at \url{http://www.raunvis.hi.is/~ath/TFBT/}}
with $Z_\mathrm{eff}$ evaluated as in Paper~II.

\begin{figure}\epsscale{1}
\plotone{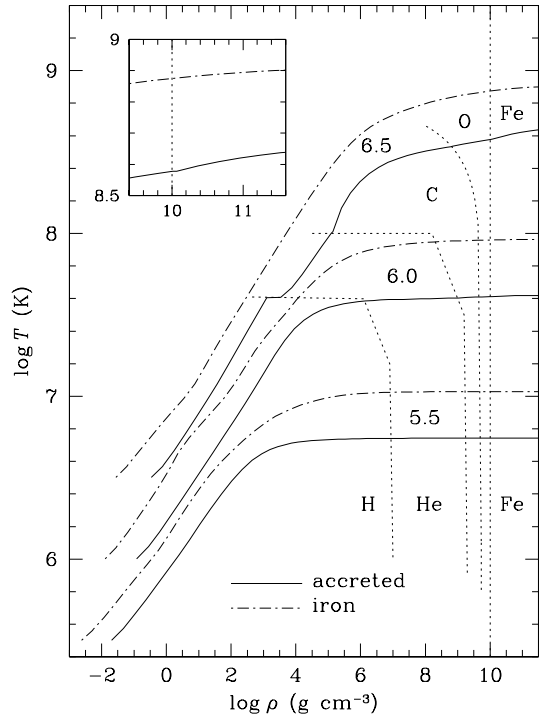}
\caption{Temperature profiles in a nonmagnetized
envelope of the canonical NS containing maximum accreted mass
(solid lines), compared with the iron envelope (dot-dashed lines).
The dotted lines correspond to interfaces between the layers of
different chemical elements. The inset demonstrates
the temperature increase in the hot envelope
behind the deepest accreted layer.
\label{fig-pt2c}}
\end{figure}

As in Paper I, the accreted envelope is modeled 
by a sequence of layers of H, He, C, O, and Fe;
and Fe is considered as the end point of nuclear
transformations in the heat-blanketing envelopes.
An example of the thermal structure of a nonmagnetic accreted
envelope is shown in Fig.~\ref{fig-pt2c}.
The boundaries
between the layers are determined by the conditions
of thermo- and pycnonuclear burning of the selected
elements and by $\Delta M$, the mass of light elements
(from H to O). These approximate boundaries are 
shown in Fig.~\ref{fig-pt2c} by dotted lines.
In that example, the envelope is \emph{fully accreted}: 
light elements reach the density
 $\sim10^{10}\gcc$, where pycnonuclear burning of oxygen
becomes efficient.
Compared to Paper~I, we have improved the position of the C/O boundary,
taking into account the results of \citet{sahrling}.
Their carbon ignition curve is approximately reproduced by
the formula
\beq
   T \approx\frac{5.2\times10^8\mbox{ K}}{
    \{1+[0.2\,\ln(\rho_\mathrm{pyc}/\rho)]^{-0.7}\}^{0.07}},
\eeq
at $\rho<\rho_\mathrm{pyc}\approx5.5\times10^9\gcc$.
The latter improvement, however, has almost no effect on $T_\mathrm{s}$.

As demonstrated recently by \citet{schatzetal01},
an explosive or steady-state burning of hydrogen
in the surface layers of accreting NSs may be strongly
affected by the rapid proton ($rp$) capture process
and extend to the elements like Sn, Sb, and Te, much heavier
than Fe. We do not consider this possibility
here but intend to analyze it in a separate publication.

\begin{figure*}\epsscale{.7}
\plotone{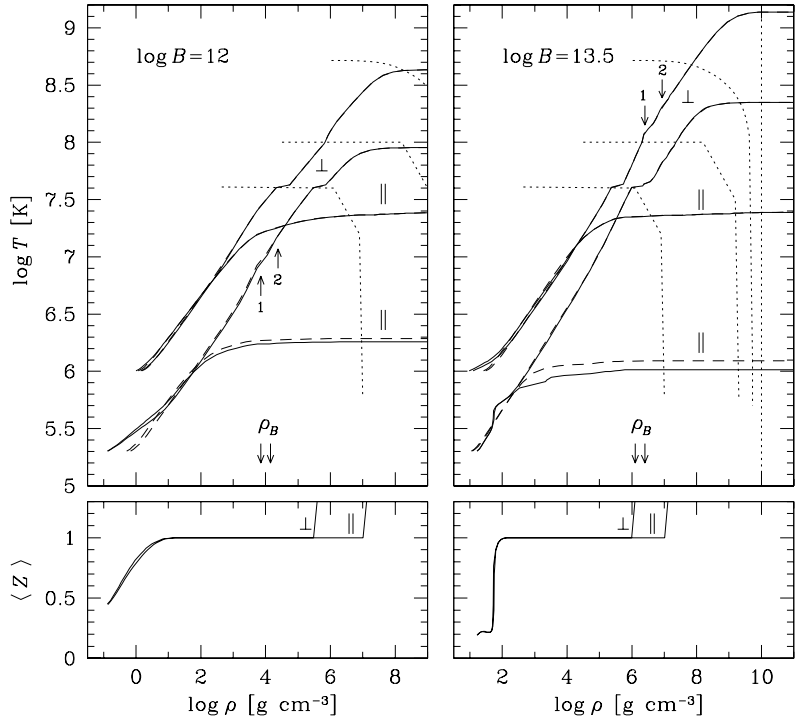}
\caption{Temperature profiles (upper panels) in the accreted
envelope ($\Delta M/M=1.5\times10^{-7}$) of the canonical NS
with $B=10^{12}$~G (left panels) and $B=10^{13.5}$~G (right), 
for $T_\mathrm{s}=2\times10^5$~K (the curves which start
in the atmosphere at lower $T$) and $T_\mathrm{s}=10^6$~K.
At each $T_\mathrm{s}$, the lower curve corresponds 
to the radial magnetic field ($\theta=0$, $\|$) and the upper to 
the tangential one ($\theta=90^\circ$, $\perp$).
The solid lines are
obtained using the hydrogen EOS
and the opacities of \citet{PC02} for the partially ionized atmosphere;
the corresponding mean ion charge number is plotted on the lower panels.
The dashed profiles are calculated using the EOS of
fully ionized ideal gas. 
The two arrows marked `$\rho_B$'
indicate the density below which the field
is strongly quantizing for hydrogen (left arrow)
or helium (right arrow on each panel). The arrows marked `1' and `2'
indicate filling of the first and the second electron
Landau levels, respectively ($\rho_B$ and $(2+\sqrt{2})\,\rho_B$,
calculated for H on the left and for $^{12}$C on the right panel).
\label{fig-po2a}}
\end{figure*}

The outermost accreted layer is assumed to consist of hydrogen. Even
a hydrogen atmosphere of a ``warm'' NS can be partially ionized in a
strong magnetic field.  Therefore we use the EOS of partially ionized
hydrogen in strong fields, derived by \citet{PCS99}  and tabulated by
\citet{PC02} at $11.9<\log_{10}B<13.5$ (where $B$ is in G). Beyond
the tabulated range we employ the model of a fully ionized ideal
electron-ion plasma. As demonstrated below (Fig.~\ref{fig-po2a}), 
the inaccuracy introduced by the latter approximation can be
significant for cold NSs ($T_\mathrm{b}<10^7$~K) with superstrong
magnetic fields  ($B>10^{13.5}$~G). The fully ionized ideal
electron-ion plasma model is used also for He, C, and O, which occupy
deeper layers.

\subsection{Opacities}
\label{sect-opac}
The heat is carried through the NS envelope
mainly by electrons at relatively high densities
and by photons near the surface. Hence,
\begin{equation}
   \kappa=\kappa_\mathrm{r}+\kappa_\mathrm{e},
\qquad
 K^{-1} = {K}_\mathrm{r}^{-1} + {K}_\mathrm{e}^{-1}, 
 \label{op}
\end{equation}      
where $\kappa_\mathrm{r}$, $\kappa_\mathrm{e}$ 
and ${K}_\mathrm{r}$, ${K}_\mathrm{e}$ denote
the radiative (r) and electron (e)
components of the thermal conductivity $\kappa$ and opacity $K$. 

Typically, the radiative conduction dominates
($\kappa_\mathrm{r}>\kappa_\mathrm{e}$) in 
the outermost nondegenerate NS layers,
whereas the electron conduction 
dominates ($\kappa_\mathrm{e}>\kappa_\mathrm{r}$)
in deeper, moderately or strongly degenerate layers.
We will see, however, that 
in the superstrong magnetic fields
the radiative conduction can be important
up to higher densities.
The $T_\mathrm{b}$--$T_\mathrm{s}$ 
relation mostly depends on the conductivities in the
\emph{sensitivity strip} on the $\rho\,$--$T$ plane
\citep{GPE}, 
where $\kappa_\mathrm{e}\sim\kappa_\mathrm{r}$.
The $T(\rho)$ profiles flatten in the inner NS zone
beyond this strip 
(e.g., see Figs.\ \ref{fig-ps2fe}, \ref{fig-pt2c}).

Following Paper~II, we use the updated electron  conductivities
$\kappa_\mathrm{e}$, presented by \citet{pbhy99} for $B=0$ and
\citet{P99} for $B\neq 0$.\footnote{Available at
\url{http://www.ioffe.ru/astro/conduct/}} Unlike the previous
expressions for $\kappa_\mathrm{e}$, the updated results take into
account multiphonon absorption and  emission processes in Coulomb
crystals and incipient long-range order in strongly coupled Coulomb
liquids of ions. As argued by \citet{baiko-ea98}, these processes are
important near the melting,  $0.3\,\Gamma_\mathrm{m} \lesssim \Gamma
\lesssim 3\Gamma_\mathrm{m}$, and they almost remove the jump of
$\kappa_\mathrm{e}$ at the liquid-solid interface.

The effect of the conductivity update on the thermal structure
at $B=0$ is illustrated
by Fig.~\ref{fig-ps2fe}. It
is marginally significant
at low $T_\mathrm{b}$ and insignificant at $T_\mathrm{b}\gtrsim10^8$~K.

The inset in Fig.~\ref{fig-pt2c} zooms the innermost 
parts of the upper (hottest) profiles for the nonaccreted and
accreted envelopes. For the accreted envelope, the
profile noticeably mounts at $\rho>10^{10}\gcc$,
because the crossing of the O/Fe interface
is accompanied by a decrease of $\kappa_\mathrm{e}$
due to the jump of $Z$.
The decrease of $\kappa_\mathrm{e}$
has smaller impact on the temperature profiles at lower $T$.

\begin{figure*}\epsscale{.7}
\plotone{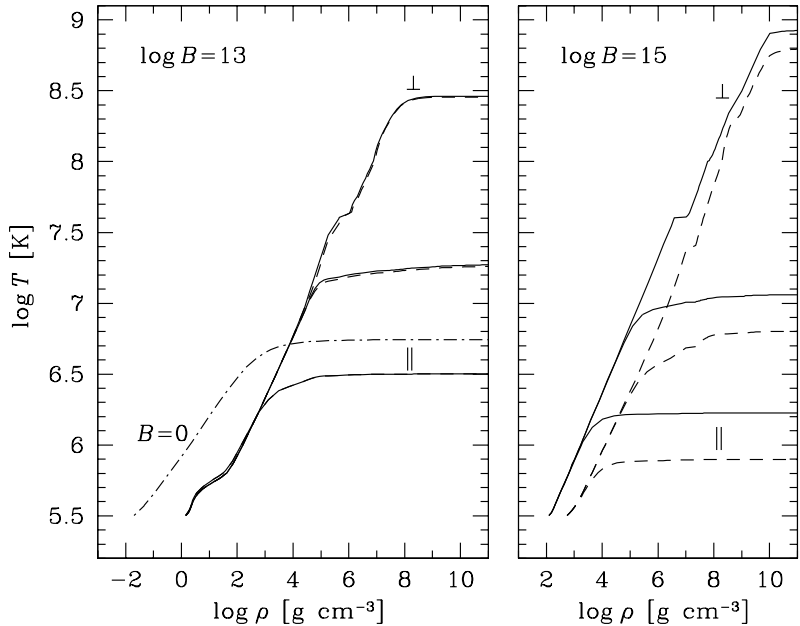}
\caption{Temperature profiles in the accreted 
envelope for the same stellar parameters as in Fig.~\ref{fig-po2a},
but at $B=10^{13}$~G (left panel) and $B=10^{15}$~G (right), 
for $T_\mathrm{s}=10^{5.5}$~K,
with (solid lines) and without (dashed lines)
plasma-frequency cut-off,
at $\theta=0$ (lower lines), $84^\circ$ (middle), 
and $90^\circ$ (upper lines). 
The dot-dashed line on the left panel is for $B=0$.
\label{fig-po2sp2}}
\end{figure*}

Radiative opacities of hydrogen at moderately strong magnetic field
are taken from \citet{PC02}.
These opacities are contributed by free-free, 
bound-free, and bound-bound transitions, and electron scattering.
They are calculated taking into account detailed ionization balance
in strong magnetic fields. 
Figure \ref{fig-po2a} shows the thermal structure of 
accreted NS envelopes at two values of $T_\mathrm{s}$ 
and two values of $B$,
for the magnetic fields radial and tangential to the NS surface.
Solid lines are obtained using the EOS and opacities of \citet{PC02},
while dashed lines correspond to the ideal fully ionized plasma model.
At $T_\mathrm{s}=10^6$~K the latter model appears to be satisfactory,
whereas at lower $T_\mathrm{s}$ there are appreciable differences.

The difference at the lowest densities (in the atmosphere)
is produced by the contribution of bound species (atoms)
in the EOS and opacities. The lower panels of Fig.~\ref{fig-po2a}
show the mean ion charge $\langle Z \rangle$ 
along the lower ($T_\mathrm{s}=2\times10^5$~K)
profiles, confirming that  $\langle Z \rangle$  
 is considerably smaller than 1 in the atmosphere.
At this relatively low temperature,
the ionization proceeds smoothly for $B=10^{12}$~G, but rather
sharply (via pressure ionization) for $B=10^{13.5}$~G.
Simultaneously with the pressure ionization, 
the temperature steeply climbs along the profiles
on the upper right panel, because 
a sharp increase of $P$ (due to the increasing number
of free particles) contributes to the right-hand side
of \req{th-str}.

The difference between the ideal-gas (dashed) and accurate 
(solid) profiles
at higher (subphotospheric) densities arises from
the Coulomb interaction.
This interaction gives a negative contribution to $P$,
thus decreasing the right-hand side of \req{th-str}.

We adopt the same radiative opacities of iron as in Papers~I and II
for zero and strong $B$, respectively.
In the first case, we use the OPAL opacities \citep{OPAL}. 
For strong $B$,
we use the free-free and scattering opacity fits from Paper~II. 
However, in the present paper (as in Paper~I
and in \citealt{PC02}) we assume that radiation
does not propagate at frequencies below the electron plasma frequency
$\omega_\mathrm{pl} =({4\pi e^2 n_\mathrm{e} / \mel} )^{1/2}$;
thus we cut off the integration 
over photon frequencies $\omega$ in the Rosseland mean opacities
at $\omega<\omega_\mathrm{pl}$.
We found that, under this assumption, the fit expressions for
$K_\mathrm{r}$ given in Paper~II should be multiplied
by a correction factor $\approx\exp\{0.005
[\ln(1+1.5\hbar\omega_\mathrm{pl}/\kB T)]^6\}$,
which effectively eliminates the radiative transport
at large densities, where
$\hbar\omega_\mathrm{pl}
     \approx 28.8\mathrm{~keV} \sqrt{\rho_6 Z/ A}\gg\kB T$.

For He, C, and O, we also employ the opacity fits from Paper~II with
the above plasma-frequency correction. The effect of this correction
on the thermal structure  of the accreted envelope is  illustrated in
Fig.~\ref{fig-po2sp2}. It is unimportant at moderately strong
magnetic fields, but quite significant at superstrong fields, because
of the high photosphere densities in the latter case.  The effect
becomes less pronounced at higher $T_\mathrm{s}$. Actually some
plasma waves  can propagate at  $\omega<\omega_\mathrm{pl}$ (e.g.,
\citealt{Ginzburg}) and carry the heat; thus we expect that realistic
temperature profiles should lie between the solid and dashed curves
in Fig.~\ref{fig-po2sp2}.

\section{Thermal structure}
\label{sect-thst}
%

\subsection{Temperature profiles}
We integrated 
Eq.~(\ref{th-str}) for various magnetic field strengths $B$,
inclination angles $\theta$,
surface temperatures $T_\mathrm{s}$, and accreted masses $\Delta M$,
using the envelope models described in the previous section.
We performed the calculations at two values of the surface gravity, 
$g=10^{14}$ cm s$^{-2}$ and $2.43\times 10^{14}$ cm s$^{-2}$,
and checked that the approximate scaling relation
$T_\mathrm{s}\propto g^{1/4}$ 
obtained by \citet{GPE} and confirmed in Papers~I and II,
holds also in the present case. 

As clear from the discussion in \S\ref{sect-input}, our EOS and
opacity models may be crude at $B \ga 10^{14}$~G. Further
improvements of physics input are required in superstrong fields.

\begin{figure}\epsscale{1.2}
\plotone{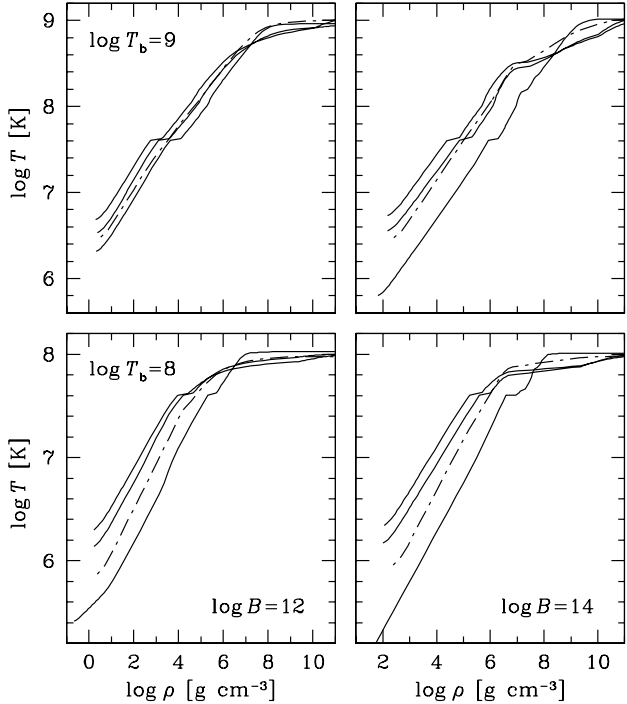}
\caption{Temperature profiles in the accreted 
envelope for the same stellar parameters as in Fig.~\ref{fig-po2a},
but at $B=10^{12}$~G (left panels) and $B=10^{14}$~G (right), 
for $T_\mathrm{b}=10^8$~K (lower panels) and 
$10^9$~K (upper ones),
at $\theta=90^\circ$ (steepest lines), $84^\circ$ (middle), and 0
 (the least steep lines). 
The dot-dashed lines correspond
to iron envelopes at $\theta=84^\circ$.
\label{fig-pj2c}}
\end{figure}

Some examples of calculated temperature profiles at several  values
of $T_\mathrm{s}$  are shown in Figs.~\ref{fig-po2a} and
\ref{fig-po2sp2}, discussed above. Figure \ref{fig-pj2c} displays the
profiles at two values of  $B$ and three values of $\theta$, for
$\Delta M=2\times10^{-7}\,M_\odot$ (solid lines) and 0 (dot-dashed
lines), and for two values of $T_\mathrm{b}$ ($10^8$ K and $10^9$ K).

\subsection{Relation between internal and effective temperatures}

Using the above results we have calculated the
$T_\mathrm{b}$--$T_\mathrm{s}$ relation
for a number of input parameters. We confirm
the conclusion of Paper I that the relation is
rather insensitive to the details of the distribution
of different chemical elements within the accreted envelope
but depends mainly on $\Delta M$, the total mass
of the elements from H to O. In particular, all
hydrogen can be replaced by He leaving the 
$T_\mathrm{b}$--$T_\mathrm{s}$ relation 
almost unchanged.

We have compared our $T_\mathrm{b}$--$T_\mathrm{s}$ relation
with that calculated by \citet{bbc02}
(their Fig.\ 4) for nonmagnetized accreted envelope
composed of H, He, and Fe. The agreement
is quite satisfactory. 

We have fitted our
numerical $T_\mathrm{b}$--$T_\mathrm{s}$ relation  
calculated for magnetized accreted envelopes
by analytic formulas (presented in the Appendix).
In the limits of $B=0$
and/or $\Delta M=0$ these formulas do not exactly reproduce the fits
obtained in Papers~I and II.
The differences reflect the improvements in the physics input
and envelope models, discussed in \S\ref{sect-input}.
For instance, in calculations shown in Fig.~\ref{fig-pj2c}, 
the boundary condition for the integration inside the star
was determined using the fitting formulas for
$T_\mathrm{s}$, presented in the Appendix.
The good convergence of the profiles toward the desired
$T_\mathrm{b}$ confirms the accuracy of our fits.

\begin{figure}\epsscale{1}
\plotone{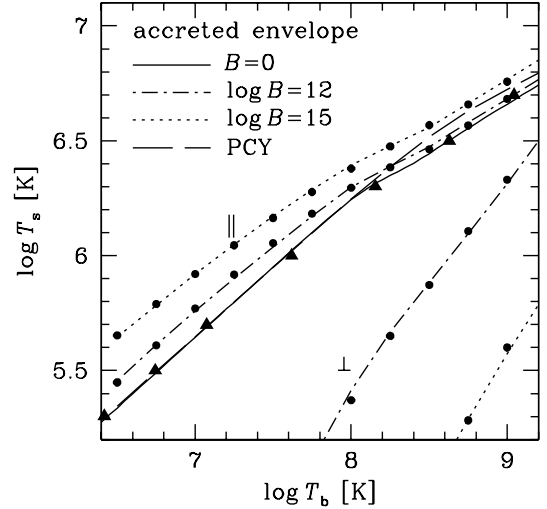}
\caption{Effective surface temperature $T_\mathrm{s}$ vs.\ temperature
at the bottom of the heat-blanketing envelope, $T_\mathrm{b}$,
for the canonical NS
with fully accreted envelope ($\Delta M=2\times10^{-7}\, M_\odot$).
Solid line: $B=0$, dot-dashed line: $B=10^{12}$ G,
dotted line: $B=10^{15}$~G, long-dashed line: 
approximation from Paper~I ($B=0$).
Heavy dots and triangles show the numerical results.
At $B\neq0$, lower lines correspond to $\theta=90^\circ$ ($\perp$) and
upper lines to $\theta=0$ ($\|$).
\label{fig-tbtsb2}}
\end{figure}

Figure \ref{fig-tbtsb2} illustrates the $T_\mathrm{b}$--$T_\mathrm{s}$
relation for the fully accreted envelope of the canonical NS
without magnetic field, with moderately strong field $B=10^{12}$ G,
and with superstrong field $B=10^{15}$ G, for two field 
geometries --- normal
($\theta=0$) and tangential ($\theta=90^\circ$) to the surface.
The lines show the fit, and the symbols (dots and triangles) show
the numerical results. The nonmagnetic fit of Paper~I
is plotted by the dashed line. Its deviation from the
present results (solid line) at high temperatures is mainly explained 
by the change of $\rho_\mathrm{b}$
(\S\ref{sect-equations}).

\begin{figure*}\epsscale{1}
\plotone{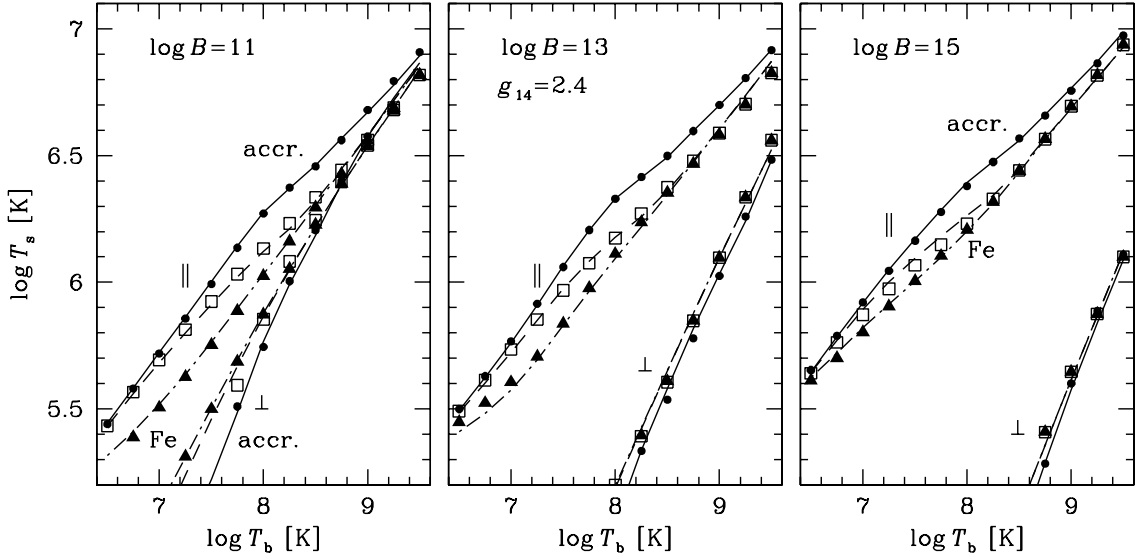}
\caption{$T_\mathrm{s}$ vs.\ $T_\mathrm{b}$
for the canonical NS
with fully accreted (solid line and dots),
iron (dot-dashed lines and triangles) 
and partly accreted envelope at $\Delta M/M=10^{-12}$
(dashed lines and empty squares), 
for $\theta=0$ (upper lines and symbols) and $\theta=90^\circ$
(lower ones).
The lines show the fit, and the symbols show the numerical results.
Left, middle, and right panels: $B=10^{11}$~G, $B=10^{13}$ G,
and $B=10^{15}$~G, respectively.
\label{fig-tb_ts_b1}}
\end{figure*}

Figure \ref{fig-tb_ts_b1} illustrates the $T_\mathrm{b}$--$T_\mathrm{s}$
relation for the canonical NS
with fully, partly, and nonaccreted envelopes 
at $\theta=0$ and $90^\circ$. 
The longitudinal heat transport ($\theta=0$) is enhanced
by the accreted envelope and/or by the strong magnetic field, 
thus increasing $T_\mathrm{s}$. However, 
the transverse transport (lower lines),
which is reduced by the strong magnetic field, 
is additionally reduced by the accreted envelope, 
further decreasing $T_\mathrm{s}$.
Actually, in longitudinal and transverse cases the lowering of $Z$  
increases the effective electron relaxation time $\tau$. 
However, at $\omg\tau\gg1$ the transverse electron conductivity
(contrary to the longitudinal conductivity or the conductivity
at $B=0$)
is \emph{inversely} proportional to $\tau$ 
(e.g., \citealt{YaK,P99}).

\subsection{Total photon luminosities}
\label{totlum}
In order to calculate the cooling curves,
one needs the total photon luminosity $L_\gamma$
as a function of $T_\mathrm{b}$. In the magnetic case,
$L_\gamma$ is given by the average over the surface, \req{L}.

\begin{figure}\epsscale{1.2}
\plotone{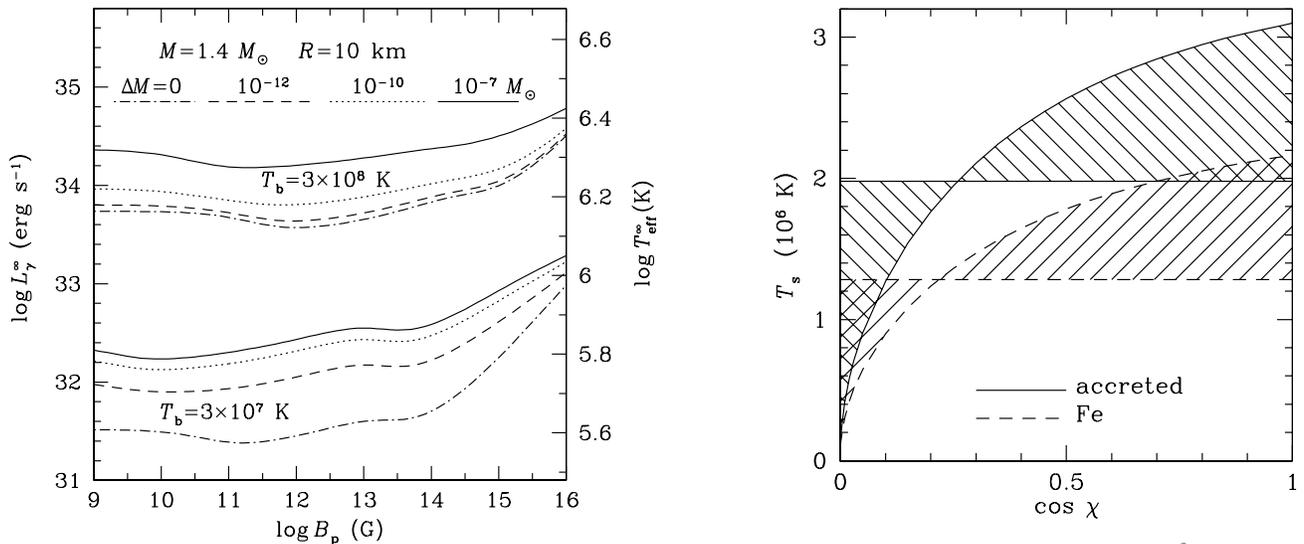}
\caption{Photon surface luminosity (redshifted as detected by a distant observer,
left vertical axis) or redshifted effective surface temperature
(right vertical axis)
of a canonical NS with a dipole magnetic field, for
two values of $T_\mathrm{b}$ and four models of the 
heat-blanketing envelope (accreted mass $\Delta M=0$,
$10^{-12} \,M_\odot$, $10^{-10} \,M_\odot$, 
or $10^{-7} \, M_\odot$)
versus magnetic field strength at the magnetic pole.
\label{fig-teb02}}
\end{figure}

\begin{figure}\epsscale{1}
\plotone{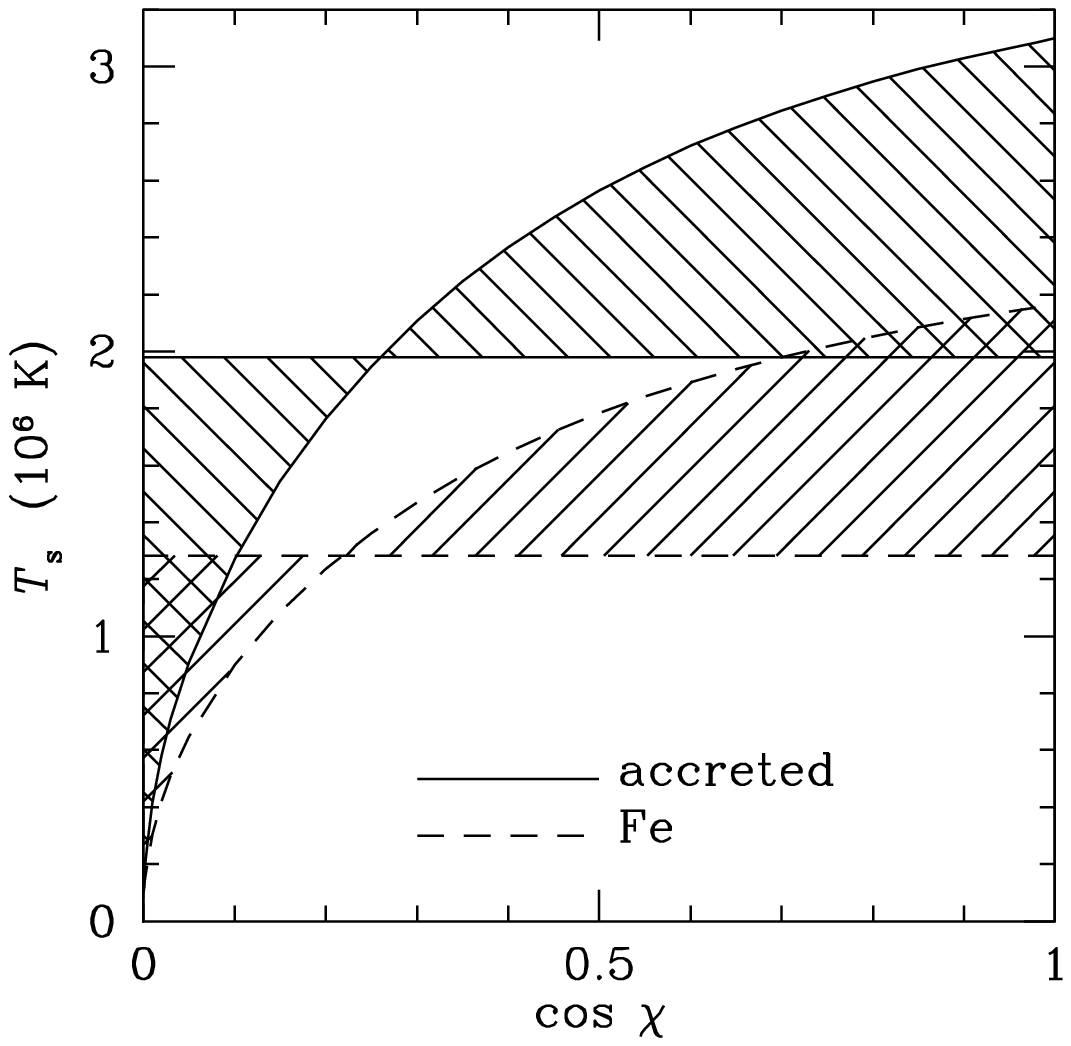}
\caption{Local surface temperature  (in $10^6$ K) vs.\ cosine
of the polar angle $\chi$ for a NS model with $M=1.4\,M_\odot$,
$R=10$ km, $T_\mathrm{b}=2\times10^8$~K, $B=0$ (horizontal lines)
and the dipole field with $B_\mathrm{p}=10^{15}$~G (curved lines),
for nonaccreted (dashed lines) and accreted (solid lines)
envelopes.
\label{fig-ts_cosl}}
\end{figure}

Figure \ref{fig-teb02} displays the photon luminosity versus $B$ for
two selected values of $T_\mathrm{b}$ ($3 \times 10^7$ and $3 \times
10^8$ K) and four selected values of $\Delta M$. The dependence of
$L_\gamma$ on $B$ is complicated.  At  $T_\mathrm{b}=3 \times 10^8$ K
(in a warm NS) and not very strong fields, the equatorial reduction
of the heat transport clearly dominates, and the NS luminosity is
lower than at $B=0$. For higher $B$, the polar enhancement of the
heat transport becomes more important, and the magnetic field
increases the photon luminosity. At $T_\mathrm{b}= 3 \times 10^7$ K (in
a much colder NS) the effect of magnetic fields $B \lesssim 10^{13}$
G is very weak while the increase of $L_\gamma$ by stronger fields is
much more pronounced. This is because the electron contribution into
conduction becomes lower (and the radiative contribution higher) in a
cold plasma. Accordingly, the equatorial decrease of the heat
transport (associated with electron conduction) is weaker. On the
other hand, the quantum effects increase both, the electron and
radiative thermal conductivities, and are more pronounced in a colder
plasma. Generally, the photon luminosity is not a simple function of
$B$, because it is affected by a number of  factors. Radiative
opacities, longitudinal electron conductivities, position of the
radiative surface
all depend on $B$ and
affect $T_\mathrm{s}$. In Fig.~\ref{fig-teb02}, the variation of
$L_\gamma$ for $B_\mathrm{p}<10^{14}$~G does not exceed a factor of
2.5.

\begin{figure*}\epsscale{1.1}
\plotone{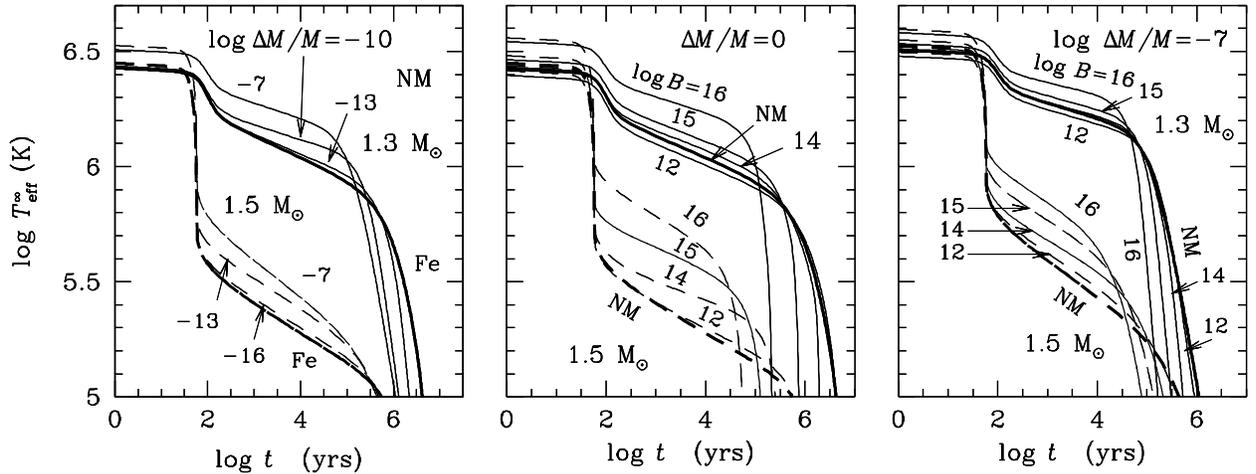}
\caption{Cooling of nonsuperfluid NSs with $M=1.3\,M_\odot$
(solid lines, slow neutrino emission) 
and $1.5\,M_\odot$ (dashed lines, fast neutrino emission) assuming 
EOS B in the NS cores. Left panel: nonmagnetic (NM) NSs with different
amount $\Delta M$ of light elements 
in the heat blanketing envelopes (the values of $\log\,\Delta M/M$
are shown near the curves); thick curves refer to
nonaccreted (Fe) envelopes. Middle panel: nonaccreted envelopes
with dipole surface magnetic fields (the curves are marked
with $\log B$, where $B=B_\mathrm{p}$ is the field at the pole);
thick lines refer to $B=0$.
Right panel: same as in the middle panel but with a fully 
accreted envelope.
\label{tript}}
\end{figure*}

A higher $B$ causes a more significant increase of the photon
luminosity. However, in this case the effect of an accreted envelope
is weaker than at $B=0$. The origin of this weakening is explained in
Fig.~\ref{fig-ts_cosl}, which illustrates the dependence of the local
effective temperature on the magnetic latitude, specified by the
polar angle $\chi$, \req{theta-chi}, for a canonical NS with
$T_\mathrm{b}=2\times10^8$~K. The dashed and solid lines refer to the
nonaccreted and fully accreted envelopes, respectively. The
horizontal lines refer to the case of $B=0$ in which the  accreted
envelope increases $T_\mathrm{eff}$ by a factor of 1.73. The curved
lines show the case  of $B_\mathrm{p}=10^{15}$~G. Since the effects
of the magnetic field weaken with increasing temperature (Paper~II),
the increase of $T_\mathrm{s}$ at the magnetic pole for the hotter
accreted envelope is smaller than the analogous increase for the
cooler iron envelope. Hence the equatorial band, where $T_\mathrm{s}$
decreases, is broader for the accreted envelope, leading to an
additional compensation of the total luminosity increase.

\section{Cooling}
\label{sect-cool}

Let us outline the effects of accreted envelopes
and surface magnetic fields on NS cooling.
We use the same nonisothermal, general relativistic
cooling code as in
\citet{gyp}, but incorporated
the above effects of accreted envelopes and
surface magnetic fields. In particular, we have shifted
the inner boundary of the heat blanketing envelope
to the neutron-drip density (see \S\ref{sect-equations}). 

\begin{deluxetable*}{cccccccc}
\tabletypesize{\small}
\tablecolumns{8} 
\tablewidth{0pc} 
 \tablecaption{Neutron star models
\label{NSmodels}}
\tablehead{
\colhead{EOS} & $M$  & $R~~~$  & $\rho_\mathrm{c}$ 
    & $M_\mathrm{crust}$ & $\Delta R_\mathrm{crust}$
    & $M_\mathrm{D}$ & $R_\mathrm{D}$  \\
    & ($M_\odot$) & (km) & ($10^{14}$ g cm$^{-3}$)   & ($M_\odot$)
    & (km)  & ($M_\odot$) & (km)}
\startdata
       & 1.3      & 13.04 &  \phn7.44 & 0.057 & 1.58 & \ldots & \ldots \\
\rb{A}        & 1.5      & 12.81 &  \phn9.00 & 0.049 & 1.26 & 0.137  & 4.27 \\
\noalign{\smallskip}\colrule\noalign{\smallskip}
        & 1.3      & 11.86 & 10.70 & 0.039 & 1.26 & \ldots & \ldots \\
\rb{B}        & 1.5      & 11.38 & 14.20 & 0.028 & 0.93 & 0.065  & 2.84 \\
\enddata
 \end{deluxetable*}

For simplicity, we assume
that the NS cores are composed of neutrons, protons, and
electrons. We adopt two model EOSs of this matter,
EOS A and EOS B, based on the
EOSs proposed by \citet*{pal88} and described, for instance, in
\citet{badhonnef}. EOS A is model I of \citet{pal88}
with the compression modulus of saturated nuclear matter
$K=240$ MeV. EOS B corresponds to $K=180$ MeV
and to the simplified form of the symmetry energy
proposed by \citet{pa92}. The NS models based
on EOSs A and B are described, for instance, by
\citet{gyp}. EOS A is somewhat stiffer, and
yields the maximum NS mass $1.98\,M_\odot$,
while EOS B yields the maximum mass of $1.73\,M_\odot$.
Both EOSs allow the powerful direct Urca process 
of neutrino emission to operate
at sufficiently high densities $\rho > \rho _\mathrm{D}$
(with $\rho_\mathrm{D}=7.85 \times 10^{14}$ g cm$^{-3}$ and
$1.298 \times 10^{15}$ g cm$^{-3}$ for EOSs A and B,
respectively). We consider the NS models
of two masses, 1.3 and 1.5 $M_\odot$.
The parameters of these models are listed in Table \ref{NSmodels}:
$\rho_\mathrm{c}$ is the NS central density,
$M_\mathrm{crust}$ the crust mass, $\Delta R_\mathrm{crust}$
the crust thickness (defined as $\Delta R_\mathrm{crust}=R-R_\mathrm{cc}$,
$R_\mathrm{cc}$ being the circumferential radius 
of the crust-core interface),
$M_\mathrm{D}$ is the mass of the inner core (if available) where direct
Urca process operates, and $R_\mathrm{D}$ is the circumferential radius
of this core.
The central densities of the low-mass NSs,
$1.3\,M_\odot$, are smaller
than $\rho_\mathrm{D}$ for both EOSs. They give an example of
slow cooling. The central densities of $1.5\,M_\odot$
NSs exceed $\rho_\mathrm{D}$, i.e., these models give us
an example of fast cooling. 

\subsection{Overall effects}
 
Figure \ref{tript} shows the effects of the surface
magnetic fields and accreted envelopes on the cooling
of 1.3 (solid lines) and 1.5 $M_\odot$
(dashed lines) nonsuperfluid NS models with EOS B
in the core.

The left panel of Fig.\ \ref{tript} illustrates the effects of
accreted envelopes in nonmagnetized NSs. We present the cooling
curves for some values of $\Delta M$, the mass of relatively light
elements (H, He, C, and O) in the heat-blanketing envelopes. The
curves for nonaccreted (Fe) envelopes are plotted by thick lines. 
The fraction of accreted mass $\Delta M/M$ varies from 0 (nonaccreted
envelopes) to $\sim 10^{-7}$ (fully accreted envelopes); a further
increase of $\Delta M$ is limited by the pycnonuclear burning of
light elements (cf.\ \citealt{PCY}).

During the first 50 years after the birth, $1.3 \, M_\odot$ and $1.5
\, M_\odot$ NSs have nearly the same surface temperatures since the
surface is thermally decoupled from the stellar interiors (e.g.,
\citealt{gyp}). Later,  after the thermalization, the direct Urca
process in the $1.5\, M_\odot$  NS makes this NS much colder. The
change of slopes of the cooling curves at $t \sim 10^5$ yr reflects
transition from the neutrino to the photon cooling stage. At the
neutrino stage, the internal stellar temperature is ruled by the
neutrino emission and is thus independent of the thermal insulation
of the blanketing envelope. The surface photon emission is determined
by the $T_\mathrm{b}$--$T_\mathrm{eff}$ relation. Since the accreted
envelopes of not too cold NSs are more heat transparent, the surface
temperature of an accreted star is noticeably higher than that of a
nonaccreted one. One can see that even a very small fraction of
accreted matter, such as $\Delta M/M \sim 10^{-13}$, can change
appreciably the thermal history of the star. The colder the star, the
smaller the fraction of accreted material which yields the same
cooling curve as the fully accreted blanketing envelope. This effect
is more pronounced for the fast cooling.

At $t \gtrsim 10^5$ yr, a star enters
the photon cooling stage.
Since the accreted blanketing envelopes
have lower thermal insulation, the NSs with such envelopes
cool faster at the photon stage than the
nonaccreted NSs (Fig.\ \ref{tript}).
Thus, the light elements make the opposite effects
on the surface temperature at the
neutrino and photon cooling stages. This \emph{reversal}
of the effect while passing from one stage to the other
is very well known and quite natural. Similar results
have been obtained in Paper~I, but our new
$T_\mathrm{eff}-T_\mathrm{b}$ relation slightly weakens the
effect of accreted envelopes.

The middle panel of Fig.\ \ref{tript} displays the effect
of dipole magnetic field on the cooling
of NSs with nonaccreted envelopes. We present the
cooling curves for several magnetic field strengths
at the magnetic poles (numbers next to the curves) up
to $B_\mathrm{p}=10^{16}$ G. The cooling curves of nonmagnetic NSs
are plotted by thick lines.   
For simplicity, the magnetic field is treated as fixed 
(nonevolved). The thermal state of the stellar interior
is almost independent of the magnetic field in the
NS envelope at the neutrino cooling stage,
but is affected by
the magnetic field later, at the photon cooling stage.
On the contrary, the surface temperature
is always affected by the magnetic field.

The dipole field $B_\mathrm{p} \lesssim 10^{13}$~G makes
the blanketing envelope of a warm ($1.3\,M_\odot$) NS
overall less heat-transparent (\S \ref{totlum}).
This lowers $T_\mathrm{eff}$ at the neutrino cooling
stage  and slows down cooling at the photon cooling stage.
The dipole field $B_\mathrm{p} \gg 10^{13}$ G makes the
blanketing envelope overall more heat transparent,
increasing $T_\mathrm{eff}$ at the neutrino cooling stage
and accelerating the cooling at the photon stage.
The field $B_\mathrm{p} \sim 10^{13}$ G has almost no effect
on the NS cooling. The presented results
are in satisfactory agreement with our previous
studies (Paper~II). 

The rapid cooling of colder magnetized (1.5 $M_\odot$) NSs
is somewhat different. 
A strong magnetic field makes the
heat-blanketing envelopes of these NSs overall more
heat transparent (\S \ref{totlum}), i.e.,
increases $T_\mathrm{eff}$
at the neutrino cooling stage and decreases it 
at the photon cooling stage. The fields $B_\mathrm{p} \lesssim 10^{13}$ G
have almost no effect on $T_\mathrm{eff}$.
On the contrary, the effects of higher
fields, $B_\mathrm{p} \gtrsim 10^{13}$ G, are much stronger than
for slowly cooling NSs.

The right panel of Fig.\ \ref{tript} presents the
cooling curves of NSs with fully accreted envelopes
and the same dipole magnetic fields as in the middle panel.
For a NS with $B_\mathrm{p} \lesssim 10^{15}$ G at the
neutrino cooling stage the effect of the accreted envelope
is stronger than the effect of the magnetic field.   
For higher $B_\mathrm{p}$, the magnetic effect dominates; the accreted
envelope produces a rather weak additional rise of $T_\mathrm{eff}$.

The main outcome of these studies is that even
ultrahigh magnetic fields
cannot change 
the average surface temperatures of young
and warm NSs as appreciably as an accreted envelope can, 
although the distribution of the local surface
temperature over the surface of a magnetized NS
can be strongly nonuniform.

\subsection{Very slowly cooling low-mass NSs}

As demonstrated recently by \citet*{ykg01,badhonnef} and
\citet{kyg02}, the effects of strong magnetic fields and accreted
envelopes are especially important in low-mass NSs (where the direct
Urca process is forbidden) with a strong proton superfluidity in
their cores. Such a superfluidity (with typical values of proton
critical temperature $T_\mathrm{cp}(\rho) \gtrsim 5 \times 10^9$ K)
fully suppresses the modified Urca process which otherwise would be
the main neutrino emission mechanism in low-mass NSs. The neutrino
emission is then generated in the reactions of neutron-neutron
scattering; it is not suppressed by a proton superfluidity and
becomes dominant.

\begin{figure}\epsscale{1}
\plotone{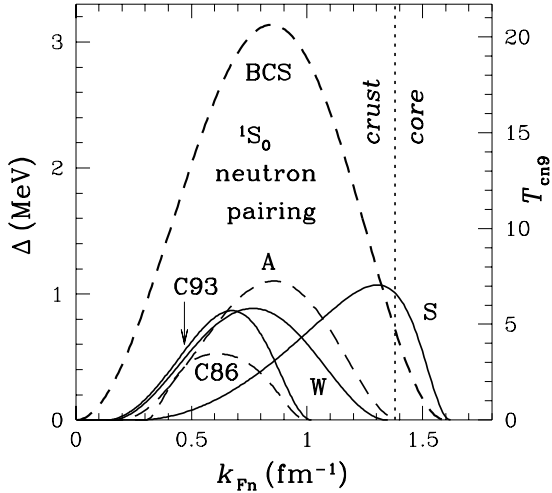}
\caption{
Energy gaps (left vertical axis) or
critical temperature (right vertical axis)
for various models of crustal neutron superfluidities 
(see the text)
versus neutron Fermi wave number; the vertical dotted
line marks the crust-core interface. 
\label{gap}}
\end{figure}

The low-mass NSs with strongly superfluid protons have thus very low
neutrino luminosity and form a special class of \emph{very
slow-cooling stars}. Their models are useful to interpret the
observations of isolated NSs hottest for their ages. According to
\citet{ykg01,ykhg02} and \citet{kyg02}, there are two NSs of such a
type among several isolated NSs whose thermal radiation has been
detected. One of them, RX J0822$-$43, is young, while the other, PSR
B1055$-$52, is much older. The third possible candidate, RX
J1856--3754, was moved (\citealt{ykhg02}) from the class of very
slow-cooling NSs to the class of faster coolers after the revision of
its age (\citealt{wl02}).     In this subsection, we focus on the
interpretation of the observations of RX J0822$-$43 and PSR
B1055$-$52, taking into account the effects of magnetic fields and
accreted envelopes. The latter effects are less important for the
interpretation of observations of colder isolated NSs (see,  e.g.,
\citealt{kyg02} and \citealt{badhonnef} for a recent comparison of
observations and cooling theories of NSs with nucleon cores).

\begin{figure*}\epsscale{1.1}
\plotone{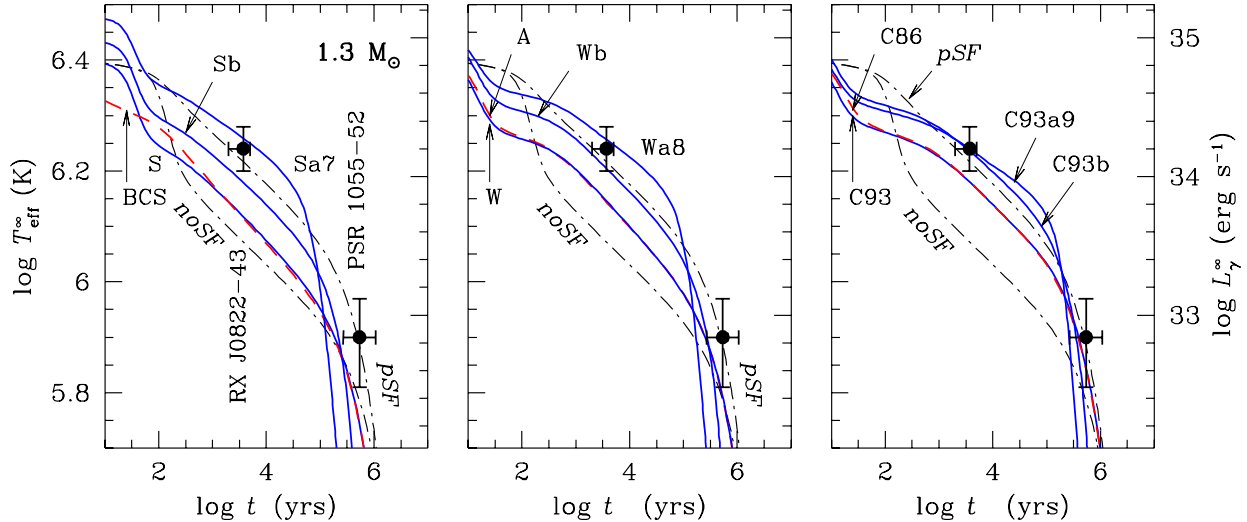}
\caption{
Evolution of redshifted effective surface 
temperature ($T^\infty_\mathrm{eff}$) or
photon luminosity ($L_\gamma^\infty$) 
of a low-mass ($1.3\,M_\odot$)
NS with EOS A confronted with
observations of RX J0822$-$43 and PSR B1055$-$52. Dot-and-dashed
curves: nonmagnetized nonaccreted
NS model without any superfluidity (\textit{noSF}) or 
with a strong proton core superfluid alone (\textit{pSF}).
Other curves are obtained
including the effects of proton core superfluidity
and a model of crustal neutron pairing from Fig.\ \ref{gap}
(BCS or S: left panel;
W or A: middle panel; C86 or C93: right panel). Models
Sa7, Wa8, and C93a9 include also the effects of
accreted envelopes with $\Delta M/M=10^{-7}$, $10^{-8}$, and
$10^{-9}$, respectively. Models Sb, Wb, and C93b
refer to nonaccreted envelopes with dipole magnetic
field $B_\mathrm{p}=10^{15}$ G. 
\label{lowm}}
\end{figure*}

Our analysis is illustrated in Figs.\ \ref{gap} and \ref{lowm}.
The observational limits $T_\mathrm{eff}^\infty=(1.6$--1.9) MK
(at the 90\% confidence level)
for RX J0822$-$43 (Fig.\ \ref{lowm}) are taken from \citet*{ztp99}.
They are obtained by fitting the observed spectrum
with the hydrogen atmosphere model; such a model is more
appropriate for this source than the blackbody model
of thermal emission. In contrast,
it seems more reasonable to fit the spectrum of PSR B1055$-$52
with a model containing a blackbody
thermal component (e.g., \citealt{pzs02}). 
The results of different groups
are not fully consistent because of
the complexity of observations
and their interpretation.
For instance, according
to G.\ G.\ Pavlov (2002, private communication)
the best-fit value obtained
from the recent \textit{Chandra} observations of PSR B1055$-$52 is
$T_\mathrm{eff}^\infty \approx 60$ eV. Attributing
a typical error bar (at the 90\% confidence level)
derived from the observations of the same source
somewhat earlier 
(G.\ G.\ Pavlov \& M.\ Teter 2002, private communication)
 we have
$T_\mathrm{eff}^\infty = (60 \pm 6)$ eV.
An analysis of the recent \textit{BeppoSAX} observations of PSR B1055$-$52
gives $T_\mathrm{eff}^\infty = (75\pm 6)$ eV
(\citealt{mineoetal02}) at the 68\% confidence level.
We adopt, somewhat arbitrarily,
a wide $T_\mathrm{eff}^\infty$ interval from
54 eV (the minimum value from the \textit{Chandra} data)
to 81 eV (the maximum value from the \textit{BeppoSAX} data).
We expect that this interval reflects the actual
uncertainty of our knowledge of $T_\mathrm{eff}^\infty$
for PSR B1055$-$52.

The NS ages are also known with some
uncertainty. For RX J0822$-$43,
we take the age range $t=2$--5 kyr (as can be deduced, e.g.,
from a discussion in \citealt*{adp91}) centered at
$t=3.7$ kyr (\citealt{wtki88}). For PSR B1055$-$52,
we adopt the standard spin-down age of 530 kyr and assume
that it is uncertain within a factor of 2.

For our analysis, we take a 1.3 $M_\odot$
NS model with EOS A in the core. We use model
1p of sufficiently strong proton superfluidity in the core (in notations
of \citealt{kyg02}) and assume a weak 
(triplet-state) neutron pairing in the core with
the maximum critical temperature lower than
$2 \times 10^8$ K. This weak neutron superfluidity
has no effect on NS cooling and can be neglected
in the cooling simulations. A stronger neutron
superfluidity in the NS core would initiate a powerful neutrino
emission due to Cooper pairing of neutrons
accelerating NS cooling in disagreement with
the observations (e.g., \citealt{ykhg02,badhonnef}).
As shown by \citet{ykg01}, the cooling of low-mass
NSs is rather insensitive to the model EOS in the NS core,
to the NS mass (as long as the mass is sufficiently
low to avoid fast neutrino cooling in the NS core),
and to the model of proton superfluidity
in the NS core (as long as the proton critical
temperature is $\gtrsim (2-3) \times 10^9$ K over the core to suppress
the modified Urca process). 
Therefore, model 1p of proton superfluid is used just as an example,
and the cooling curves are actually independent of the 
features of strong
proton superfluid. 
Let us
add that it is likely that the cores of
low-mass NSs consist of nucleons and electrons
(with forbidden direct Urca process). 
Muons may also be present there, but 
have almost no effect on the cooling of low-mass NSs \citep{BejgerYak}.
Thus,
our cooling scenarios are not related to the
models of matter of essentially supranuclear
density in the inner cores of massive NSs
where the composition may be exotic (e.g., includes pion or kaon
condensates, or quark matter).

We see that the models of low-mass cooling NSs are
sufficiently robust against uncertainties in the physics
of matter in the NS cores. Accordingly,
the cooling of this special class of NSs
(contrary to the cooling of other NS models)
is \emph{especially sensitive to the
properties of the NS crust}.
It is mainly regulated by the effects of (\textit{i})
accreted matter and (\textit{ii}) surface magnetic fields
in the heat-blanketing envelopes, as well as by the
effects of (\textit{iii}) singlet-state neutron superfluidity
in the inner NS crusts. All these effects are
of comparable strength. They are analyzed below
and illustrated in Figs.\ \ref{gap} and \ref{lowm} with the aim
to interpret the observations of RX J0822$-$43
and PSR B1055$-$52.

Two dot-and-dashed curves marked as \textit{noSF} and \textit{pSF}
in each panel of Fig.\ \ref{lowm} show the cooling of
a nonsuperfluid NS and a NS with strong proton superfluidity
in the core (the effects of magnetic fields
and accreted envelopes are neglected). 
We see that the proton superfluidity, indeed, delays
the cooling (by suppressing the modified Urca process)
and makes the cooling curves consistent with
the observations of RX J0822$-$43 and PSR B1055$-$52 \citep*{khy01}.

However, our ``successful'' \textit{pSF} cooling curve
neglects the presence of neutron
superfluidity in the NS crust. This superfluidity is
predicted by microscopic theories (e.g., \citealt{ls01})
although the superfluid gaps are very model dependent.
The superfluidity initiates neutrino emission due to Cooper pairing
of crustal neutrons which may noticeably accelerate
the cooling of low-mass NSs (\citealt{ykg01}).
In order to illustrate this effect, we consider several
models of the crustal superfluidity. 
Figure \ref{gap} shows the dependence of the superfluid
gap $\Delta$ (left vertical axis) or
associated critical temperature $T_\mathrm{cn}$
(in units of $10^9$ K, right vertical axis)
on neutron Fermi wave number $k_\mathrm{Fn}$ (as a measure of density)
for six models (from \citealt{ls01}). Vertical
dotted line indicates approximate position of
the crust-core interface (assumed to be at
$1.5 \times 10^{14}$ g cm$^{-3}$ in our cooling models).

Model BCS 
is the basic model of singlet-state neutron
pairing calculated under the simplified assumption
of purely in-vacuum neutron-neutron interaction (neglecting
medium polarization effects).
Other five models --
C86 (\citealt{C86}),   
C93 (\citealt{C93}),   
A   (\citealt*{A}),     
W   (\citealt*{W}),     
and 
S   (\citealt{S}) -- 
include medium polarization effects which weaken the strength
of neutron pairing.

In Fig.\ \ref{lowm}, in addition to the \textit{noSF} and \textit{pSF}
curves, we show six cooling curves (BCS, S, A, W, C86, and C93)
calculated adopting proton superfluidity in the NS core and one of
the models of neutron superfluidity in the crust from Fig.\ \ref{gap}
(neglecting surface magnetic fields and accreted envelopes). One can
see that the crustal superfluidity, indeed, accelerates the cooling
and complicates the interpretation of the observations of RX
J0822$-$43 and PSR B1055$-$52. The new six curves are naturally
divided into three pairs shown in three panels of Fig.\ \ref{lowm}
(left panel: BCS and S; middle panel: A and W; right panel C86 and
C93). The curves within each pair are very close, while the pairs
differ from one another. As clear from Fig.\ \ref{gap}, the neutron
superfluid gaps for any pair, although different, have one common
property: the same maximum density of superfluidity disappearance.
This maximum density limits the density of the outer NS layer where
the neutrino emission due to the singlet-state pairing of neutrons
operates and accelerates the cooling. For instance, neutron
superfluids BCS and S extend to higher densities (penetrate into the
NS core); Cooper pairing neutrino emission is generated from an
extended layer and produces the most dramatic effect: the cooling
curves go much lower than the \textit{pSF} curve, strongly violating the
interpretation of RX J0822$-$43 and almost violating the
interpretation of PSR B1055$-$52. Superfluids A and W die at lower
$\rho$ (at the crust-core interface), Cooper-pairing neutrinos are
emitted from a smaller volume, and the cooling curves go higher,
closer to the \textit{pSF} curve. They do not explain RX J0822$-$43 but
seem to explain PSR B1055$-$52. Finally, superfluids C86 and C93
disappear even at lower $\rho$, far before the crust-core interface;
the emission volume gets smaller, and the cooling curves shift even
closer to the \textit{pSF} curve simplifying the interpretation of RX
J0822$-$43 and easily explaining PSR B1055$-$52.

All NSs should have the same EOS and superfluids
in their interiors but may have different magnetic fields
and accreted envelopes. The effects of latter factors are
also illustrated in Fig.\ \ref{lowm}.
As discussed above, the cooling histories
of NSs with the superfluid model
BCS are almost the same as with S, with model A are the same as with
W, and with model C86 are the same as with C93. Therefore,
we do not consider the effect of magnetic fields and
accreted envelopes on the models C86, A and BCS.
Let us remind, that magnetic fields and accreted envelopes
have opposite effects on the thermal states of NSs at the neutrino
and photon cooling stages. PSR B1055$-$52
is just passing from one cooling stage to the other
and has no
superstrong magnetic field. It is not expected to possess an
extended accreted envelope. Thus, the effects of magnetic fields
and accreted envelopes on the evolution of this pulsar are
thought to be minor. 

The superfluid model S (or BCS) without any magnetic field
or accreted envelope is 
only marginally consistent with the observations
of PSR B1055$-$52 (left panel of Fig.\ \ref{lowm}). 
If, however, we accept this model,
then we can explain RX J0822$-$43 by switching on
the effects of the accreted envelopes or surface magnetic fields.
Curve Sb on the left panel of Fig.\ \ref{lowm} is calculated adopting
proton superfluid in the NS core,
 superfluid S in the crust, and the dipole magnetic
field with $B_\mathrm{p}=10^{15}$ G (the magnetar hypothesis).
Curve Sa7 is obtained for the same crustal superfluid, but for
$B=0$ and for $\Delta M/M=10^{-7}$ of accreted material
on the NS surface. We see that both curves, Sb and
Sa7, are consistent with the observations of
RX J0822$-$43. The magnetic field slightly above
$B_\mathrm{p}=10^{15}$ G would further improve
the agreement of the theory and observations. 

The next superfluid model W (or A) is
acceptable to interpret the observations of PSR B1055$-$52 
(Fig.\ \ref{lowm}, middle panel).
We can adopt it and switch on the
effects of the magnetic field or accreted envelope
to interpret RX J0822$-$43. Curve Wb on the right panel
corresponds to the
crustal superfluid W and the dipole magnetic
field with $B_\mathrm{p}=10^{15}$ G.
Curve Wa8 is calculated for the same crustal superfluid,
$B=0$, but for $\Delta M/M=10^{-8}$.
Both curves, Wb and
Wa8, are seen to be consistent with the observations of
RX J0822$-$43.

Finally, model C93 (or C86) of crustal superfluid is
in reasonable agreement with the data on PSR B1055$-$52
(Fig.\ \ref{lowm}, right panel).
Adding the effect of the surface magnetic field
($B_\mathrm{p}=10^{15}$ G, curve C93b) or the
accreted envelope ($\Delta M/M=10^{-9}$, curve C93a9)
we can easily explain the observations of RX J0822$-$43.

Thus all the models of crustal superfluidity are currently consistent
with the observations (although models BCS and S seem to be less
likely for explaining PSR B1055$-$52). This is a consequence of wide
observational error bars of $T_\mathrm{eff}^\infty$ for PSR B1055$-$52
(see above). We expect to constrain the models of crustal
superfluidity after better determination of $T_\mathrm{eff}^\infty$
in the future observations of PSR B1055$-$52. Afterwards it will be
possible to constrain the surface magnetic fields and the mass of the
accreted envelope from the observations of RX J0822$-$43.

It is clear from Figs.\ \ref{tript} and \ref{lowm} that old and warm
NSs like PSR B1055$-$52  cannot possess bulky accretion envelopes
which would operate as efficient coolers. This conclusion is in line
with the observations of thermal radiation from old sources: their
spectra are better fitted (e.g., \citealt{pzs02}) with the blackbody
model of thermal emission (suitable for nonaccreted matter) than with
the hydrogen atmosphere models. This conclusion is also in line with
theoretical studies of \citet{cb02} who show that light elements can
be burnt in old ($t \gtrsim 10^5$ yr) and warm NSs by diffusive
nuclear burning. In this connection, it would be interesting to
construct the models of cooling NSs incorporating the effects of
diffusive burning and associated thinning of the  light-element
envelopes in time. Cooling of magnetars  is expected to be accompanied
by magnetic field decay which should also be taken into account in
advanced cooling simulations.

\section{Conclusions}
\label{sect-concl}

We have studied the thermal structure of a heat-blanketing envelope
of a NS with a strong magnetic field and arbitrary amount of
light-element (accreted) material. We have calculated and fitted by
an analytic expression the relation between the neutron-star internal
and surface temperatures in a local element of the heat-blanketing
envelope as a function of magnetic field strength and geometry, the
mass of accreted material, and the surface gravity. We have performed
numerical simulations of cooling of NSs with dipole magnetic fields
and accreted envelopes. We have considered slow and fast cooling
regimes but mainly focused on a very slow cooling of low-mass NSs
with strong proton superfluidity in their cores. These NSs form a
special class of NSs whose thermal history is rather insensitive to
the physics of matter in the stellar cores but is mainly regulated by
the magnetic field strength and the amount of accreted material in
the heat-blanketing envelopes, as well as by the singlet-state
pairing of neutrons in the inner stellar crusts. We show that these
cooling regulators are important for explaining the observations of
thermal radiation from isolated NSs warmest for their ages, RX
J0822$-$43 and PSR B1055$-$52. Our analysis indicates that all
realistic microscopic models of the singlet-state neutron pairing are
currently consistent with the observations of PSR B1055$-$52.
Adopting these models and tuning the strength of the magnetic field
and/or the mass of accreted material we can explain also the
observations of RX J0822$-$43. We expect that such an interpretation
will be refined after new, high-quality observations of RX J0822$-$43
and PSR B1055$-$52 will appear or new very slowly cooling NSs will be
discovered. In particular, this will allow one to discriminate
between the models of crustal superfluidity and determine the depth
of superfluidity disappearance in high-density matter. It will also
be possible to constrain the surface magnetic fields and the mass of
accreted envelopes of these objects. Our results will also be useful
for constructing advanced models of cooling NSs taking into account
the evolution of strong surface magnetic fields and the mass of
light-element envelopes (e.g., under the action of diffusive nuclear
burning).

We believe that the $T_\mathrm{b}$--$T_\mathrm{s}$ relation obtained
in this paper is reliable at $B\lesssim10^{14}$ G. The results
presented for higher fields are rather indicative, because
the dense plasma effects on the heat conduction by photons near the
bottom of NS photosphere at the superstrong fields have not yet been
explored in detail.

\acknowledgements
We thank Forrest Rogers and Carlos Iglesias for providing the
monochromatic opacities and EOS for iron at $B=0$. 
We also thank the referee for useful comments.
D.Y. is indebted
to Victor Khodel for the idea to use the models
of neutron superfluidity presented in Fig.\ \ref{gap}
and to George Pavlov for providing the
data on PSR B1055$-$52. 
The work of A.P. and D.Y. was supported in part by RFBR grants
02-02-17668 and 03-07-90200.

\begin{appendix}

\section{Fitting formulas 
for the $\uppercase{T}_\mathrm{b}$--$\uppercase{T}_\mathrm{s}$ relation}
\label{sect-fit}

Let $T_\mathrm{b9}=T_\mathrm{b}/10^9$ K,
$T_\mathrm{s6}=T_\mathrm{s}/10^6$ K, and $\eta=g_{14}^2 \Delta M/M$,
where $g_{14} = g/10^{14}$ cm s$^{-2}$ and $\Delta M$ is the mass of
accreted light elements from H to O. We assume that $\eta<10^{-6}$;
at larger $\eta$ the light elements would undergo efficient
pycnonuclear burning. The envelope is fully accreted if
$\eta=10^{-6}$, and nonaccreted if $\eta=0$. Let $T_\mathrm{a6}$ and
$T_\mathrm{Fe6}$ denote  $T_\mathrm{s6}$ for the fully accreted  and
nonaccreted envelopes, respectively.

First consider the case of $B=0$.
In this case, the fit given by Eqs.~(A6)
and (A7) of Paper~I for the nonaccreted envelope remains valid:
\beq
\label{TFe}
   T_\mathrm{Fe6}^4 = g_{14}\,[(7\zeta)^{2.25}+(0.33\zeta)^{1.25}],
\quad
\zeta = T_\mathrm{b9} - 10^{-3}\, g_{14}^{1/4}\,\sqrt{7T_\mathrm{b9}}.
\eeq
For the fully accreted envelope, we have
\beq
   T_\mathrm{a6}^4 = \Big[ g_{14}\, (18.1\, T_\mathrm{b9})^{2.42} \,
   \big\{ 0.447+0.075\,(\log_{10} T_\mathrm{b})
    / [1+(6.2\, T_\mathrm{b9})^4 ]
   \big\} + 
   3.2\, T_\mathrm{b9}^{1.67}\, T_\mathrm{Fe6}^4 \Big]
  /
   (1+3.2\, T_\mathrm{b9}^{1.67} ).
\label{Ta2}
\eeq
This fit differs from the equations in Appendix C of Paper~I by 
the correction factor in curly brackets.
The difference is most appreciable
at $T_\mathrm{b}>10^8$~K,
as explained in \S\ref{sect-equations} (cf.\ Fig.~\ref{fig-tbtsb2}).

Next consider the fully accreted and nonaccreted 
magnetized envelopes. In both cases, we can write
\beq
   T_\mathrm{Fe6,a6}(B)=T_\mathrm{Fe6,a6}(0)
   \mathcal{X},
\eeq
where the magnetic correction factor $\mathcal{X}$ depends on 
$B$, $T_\mathrm{b}$, and $\theta$. It can be fitted by
\bea
   \mathcal{X} &=& \left(\mathcal{X}_\|^\alpha \cos^2\theta
   + \mathcal{X}_\perp^\alpha \sin^2\theta\right)^{1/\alpha},
\quad
   \alpha = \cases{
   4+\sqrt{\mathcal{X}_\perp/\mathcal{X}_\|}
    & for $T_\mathrm{Fe6}$, 
                           \cr\noalign{\smallskip}
         ( 2 +  \mathcal{X}_\perp/\mathcal{X}_\|)^2
                 & for $T_\mathrm{a6}$,\cr}
\\
   \mathcal{X}_\| &=& \left( 1 + \frac{a_1+a_2\,T_\mathrm{b9}^{a_3}
                }{
                T_\mathrm{b9}^2 + a_4 T_\mathrm{b9}^{a_5} }\,
                \frac{B_{12}^{a_6}
                   }{
                   (1+a_7 B_{12} 
                   / T_\mathrm{b9}^{a_8})_{\phantom{b9}}^{a_9} } \right)
       \left( 1+\frac{1}{3.7+ (a_{10}+a_{11}\,B_{12}^{-3/2})\,
          T_\mathrm{b9}^2  }  \right)^{-1},
\label{Xpar}\\
   \mathcal{X}_\perp &=& 
    \frac{\left[ 1 + b_1\,B_{12}/(1+b_2\,T_\mathrm{b9}^{b_7}) \right]^{1/2}
       }{
 \left[ 1 + b_3\,B_{12}/(1+b_4\,T_\mathrm{b9}^{b_8}) \right]^{\beta\phantom{/}}
       } ,
\quad
   \beta=\left( 1+b_5\,T_\mathrm{b9}^{b_6} \right)^{-1},
\label{Xperp}
\eea
with the parameters $a_i$ and $b_i$ given in Table~\ref{table}.

Finally, the surface temperature of a partly accreted envelope
is approximately reproduced by the interpolation:
\begin{mathletters}
\bea
   T_\mathrm{s} &=& \left[ \gamma\,T_\mathrm{a6}^4 + (1-\gamma)\,T_\mathrm{Fe6}^4 \right]^{1/4},
  \\
\gamma &=& \left[ 1 + 3.8\,(0.1\xi)^9 \right]^{-1}
        \left[ 1 + 0.171\,\xi^{7/2}\,T_\mathrm{b9} \right]^{-1} ,
\quad
   \xi = -\log_{10}(10^6\eta).
\eea
\end{mathletters}

These fitting formulas have been checked against
calculations for input parameters
restricted by the conditions $6.5 < \log_{10}T_\mathrm{b}/\mbox{K}<9.5$,
$\log_{10}T_\mathrm{s}/\mbox{K}>5.3$, and $10<\log_{10}B/\mbox{G}<16$. 
The numerical values of $T_\mathrm{s}$
are reproduced
with root-mean-square residuals of (3--5)\% ($<2$\% for fully accreted
envelopes at $\theta=0$) and maximum deviations within $\approx20$\%
(within 13\% for $\theta=0$, 
within 12\% for fully accreted or nonaccreted envelopes
at any $\theta$, 
and within 5.2\% for fully accreted envelopes at $\theta=0$).

We emphasize that our numerical results, and hence the fitting
formulas given here, are uncertain at superstrong fields ($B>10^{14}$
G) because of the dense plasma effects discussed in \$\ref{sect-opac}
(see the right panel of Fig.~\ref{fig-po2sp2}). A thorough study of
the radiative heat conduction at $\omega<\omega_\mathrm{pl}$ is
needed to obtain a reliable $T_\mathrm{b}$--$T_\mathrm{s}$ relation at
such field strengths.
\end{appendix}

\begin{deluxetable*}{ccccccccccccc}
\tabletypesize{\small}
\tablecolumns{13} 
\tablewidth{0pc}  \tablecaption{Parameters of Eqs.~(\ref{Xpar}), (\ref{Xperp})
 \label{table}}
\tablehead{envelope & $n$ & {1} & {2} & {3} & {4} & {5} & {6} & {7} & {8} & {9} & {10} & {11} }
\startdata
 & $a_n$ & 1.76E$-$4 & 0.038 & 1.5 & 0.0132 & 0.620 & 0.318 & 2.3E$-$9 & 3 & 0.160 & 21 & 4.7E+5 \\
\rb{iron} & $b_n$ & 159 & 270 & 172 & 110 & 0.363 & 0.181 & 0.50 & 0.619 \\
\noalign{\smallskip}\colrule\noalign{\smallskip}
 & $a_n$ & 4.50E$-$3 & 0.055 & 2.0 & 0.0595 & 0.328 & 0.237 & 6.8E$-$7 & 2 & 0.113 & 163 & 3.4E+5 \\
\rb{accreted} & $b_n$ & 172 & 155 & 383 & 94 & 0.383 & 0.367 & 2.28 & 1.690 \\
 \enddata
 \end{deluxetable*}


\end{document}